\documentclass[10pt,twocolumn,letterpaper]{article}

\usepackage{wacv}

\usepackage{graphicx}
\usepackage{amsmath}
\usepackage{amssymb}
\usepackage{booktabs}
\usepackage{url}            
\usepackage{amsfonts}       
\usepackage{nicefrac}       
\usepackage{microtype}      
\usepackage{multirow}
\usepackage{multicol}
\usepackage{wrapfig}
\usepackage{booktabs} 
\usepackage{makecell}
\usepackage{amsmath}
\usepackage{amssymb}
\usepackage{comment}
\usepackage{color}
\usepackage{amsmath}
\usepackage{amssymb}
\usepackage{booktabs}
\usepackage{algorithm}
\usepackage{algpseudocode}
\usepackage{wrapfig}
\usepackage{lipsum}
\usepackage{wrapfig}
\usepackage{mathtools}
\usepackage{bm}
\usepackage{enumitem}
\usepackage{bbding}
\usepackage{bbm}
\usepackage{caption}
\usepackage{varwidth}
\usepackage[export]{adjustbox}
\usepackage[dvipsnames]{xcolor}
\usepackage{xspace}
\usepackage[pagebackref,breaklinks,colorlinks]{hyperref}
\usepackage{tikz}
\usepackage{pgfplots}
\newcommand{\name}{SoundLoc3D\xspace}
\usepackage{subcaption}

\hypersetup{
linkcolor=BrickRed
,citecolor=Green
,filecolor=Mulberry
,urlcolor=NavyBlue
,menucolor=BrickRed
,runcolor=Mulberry
,linkbordercolor=BrickRed
,citebordercolor=Green
,filebordercolor=Mulberry
,urlbordercolor=NavyBlue
,menubordercolor=BrickRed
,runbordercolor=Mulberry
}

\definecolor{topcolor}{rgb}{1,0.8,0.8}
\definecolor{secondcolor}{rgb}{1,0.87,0.7}
\definecolor{thirdcolor}{rgb}{1,1,0.8}
\usepackage{colortbl}
\usepackage{fontawesome}

\DeclareMathAlphabet{\pazocal}{OMS}{zplm}{m}{n}

\DeclareMathAlphabet\mathbfcal{OMS}{cmsy}{b}{n}

\usepackage[capitalize]{cleveref}
\crefname{section}{Sec.}{Secs.}
\Crefname{section}{Section}{Sections}
\Crefname{table}{Table}{Tables}
\crefname{table}{Tab.}{Tabs.}

\begin{document}

\title{SoundLoc3D: Invisible 3D Sound Source Localization and Classification Using a \\ Multimodal RGB-D Acoustic Camera}

\author{Yuhang He\footnotemark[2]\ \ \thanks{The work was done while interning at MERL.} \hspace{0.2cm} Sangyun Shin\footnotemark[2] \hspace{0.2cm} Anoop Cherian\footnotemark[3] \hspace{0.2cm} Niki Trigoni\footnotemark[2] \hspace{0.2cm} Andrew Markham\footnotemark[2]\\
\textbf{\footnotemark[2]}\hspace{0.2cm} Department of Computer Science, University of Oxford, UK.\\
\textbf{\footnotemark[3]}\hspace{0.2cm} Mitsubishi Electric Research Labs, Cambridge, MA, US.\\
\faEnvelopeO\ \texttt{yuhang.he@cs.ox.ac.uk}
}
\maketitle

\begin{abstract}
    Accurately localizing 3D sound sources and estimating their semantic labels -- where the sources may not be visible, but are assumed to lie on the physical surface of objects in the scene -- have many real applications, including detecting gas leak and machinery malfunction. The audio-visual weak-correlation in such setting poses new challenges in deriving innovative methods to answer \textit{if} or \textit{how} we can use cross-modal information to solve the task. Towards this end, we propose to use an acoustic-camera rig consisting of a pinhole RGB-D camera and a coplanar four-channel microphone array~(Mic-Array). By using this rig to record audio-visual signals from multiviews, we can use the cross-modal cues to estimate the sound sources 3D locations. Specifically, our framework~\emph{\name} treats the task as a \textit{set prediction} problem, each element in the set corresponds to a potential sound source. Given the audio-visual weak-correlation, the set representation is initially learned from a single view microphone array signal, and then refined by actively incorporating physical surface cues revealed from multiview RGB-D images. We demonstrate the efficiency and superiority of \emph{\name} on large-scale simulated dataset, and further show its robustness to RGB-D measurement inaccuracy and ambient noise interference.
    \vspace{-4mm}
\end{abstract} 
\vspace{-1mm}
\section{Introduction}
\vspace{-2mm}
The task of 3D sound source localization and classification, which aims at localizing the 3D spatial position of each sound source~(either physical position or the direction of arrival~(DoA)) and inferring its semantic label~(\eg, \emph{a telephone ring}), has numerous applications in a variety of real-world scenarios, including robotics~\cite{soundscape_dataset,seldnet}, audio surveillance~\cite{audio_assesment,robust_audio_surv}, smart homes ~\cite{smart_room,video_conference}, and augmented/virtual reality~(AR/VR)~\cite{3Dimmersive,immersive_spatialaudio}. Existing methods for 3D sound source localization and classification can be divided into two main types: 1) methods that are vision-agnostic~\cite{sounddet,ein_v2,seld_dcase19,sounddoa} and thus rely solely on acoustic signals, and 2)  methods that use the synergy between acoustic and visual modalities by assuming that the sound source is visually discriminative~\cite{proposal_based_paradigm,visual_guide_sss,mo2023oneavm,mo2023audiovisual,senocak2018learning}. For example, the sound is heard from a musical instrument that is visually observable.

\begin{figure}[t]
    \centering
    \vspace{-3mm}    
    \includegraphics[width=0.49\textwidth]{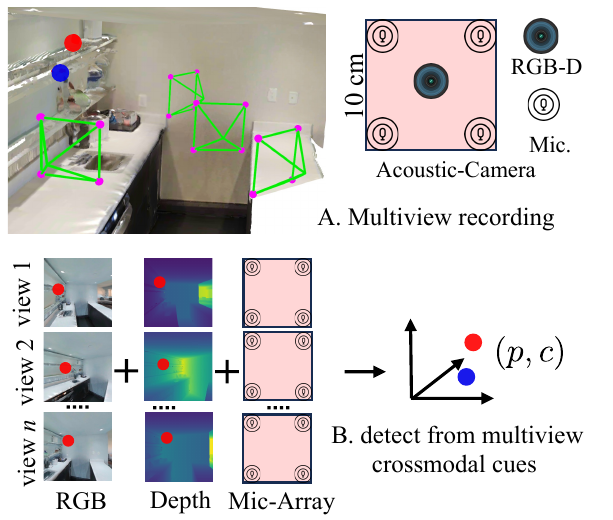}
    \caption{\textbf{SoundLoc3D problem setup}: Visually invisible sound sources freely lie on physical object's surface and are emitting sound, A: We use an acoustic-camera to record Mic-Array signal and RGB-D images from multiview. \textit{SoundLoc3D} incorporates multiview crossmodal RGB images, depth maps and Mic-Array signal to jointly localize source position~$p$ and semantic label~$c$.}    
    \label{fig:teasing_img1}
    \vspace{-4mm}
\end{figure}

While these two settings have been well-explored by audio/speech processing and audio-visual multimodal research communities, respectively, there is a third important setting that is currently under-explored, but reflects many real-world scenarios: when the sound and vision are weakly correlated; for example, when the source is either too small to be visually observed,  blended with background in texture and appearance, or has no associated visual form at all. Typical examples include: {gas leak from pipes}, {water dripping}, {electrical zapping}, {abnormal sounds of cooling fans in computers}, etc., among others. In many cases, the task of accurately detecting the source of such sounds and inferring their semantic labels is important for damage pre-diagnosis.

\setlength{\intextsep}{10pt}

Tackling the above task naturally poses three research questions: 1) is cross-modal information useful in a sound-vision weak-correlation setting? and if so 2) what sort of visual cues can be used?, and 3) how to effectively use cross-modal cues within the weak-correlation setting? In the pioneering work of Sound3DVDet~\cite{He_2024_WACV}, the focus is primarily in using cross-modal multiview RGB images towards addressing~Q1 and Q2. In this paper, we go further to explore -- in addition to multiview RGB images -- whether the depth map of the scene can be used to further improve the performance. The motivation for incorporating the depth maps is two-fold. First, alongside the RGB images, the depth maps can be easily collected~(up to an accuracy rate) using either direct depth sensors or stereo matching~\cite{Zhao_2023_CVPR_stereovision}. Second, the depth map provides more direct cue of the object's physical surface than RGB images. The integration of RGB images and depth maps thus could benefit the task by leveraging the well-explored vision-based multiview geometry methods~(\eg, S\textit{f}M~\cite{schoenberger2016sfm} and feature matching~\cite{sun2021loftr,liu2010sift}). Thus, the core questions we seek to answer are: 1) \textit{how to} incorporate multiview RGB-D images to solve the 3D sound source localization problem within the audio-visual weak-correlation setting? and 2) \textit{how to} design a framework that is robust to RGB-D measurement inaccuracy and ambient noise interference? Fig.~\ref{fig:teasing_img1} shows the problem setup.

To solve this task, we propose \emph{\name} -- an effective, unified and scalable framework for visually invisible 3D sound source localization and classification. To ensure \emph{\name} is scalable to handle arbitrary 3D sound sources, we follow~\cite{He_2024_WACV} to treat this task as a \textit{set prediction} problem~\cite{detr3d,DETR}, each element in the set is a \textit{query} and associated with a potential sound source. Given the audio-visual weak-correlation, \emph{\name} learns an initial set representation from each of the single-view Mic-Array signal and subsequently optimizes the set representation by actively incorporating sound source cues revealed by multiview RGB-D images and crossview estimation consistency revealed from multiview observations. Specifically, using the cross-modal RGB-D images, we constrain the sound source to lie on object's physical surface by encouraging: 1) \textbf{visual appearance consistency} from multiview RGB images in a feature space, and 2) \textbf{spatial proximity} of the source informed by multiview depth maps. From the cross-view observation perspective, we encourage 3) \textbf{cross-view estimation consistency} of 3D sound source.

To evaluate \emph{\name}, we run experiment on large-scale simulated multiview RGB-D and Mic-Array dataset by following Sound3DVDet~\cite{He_2024_WACV} data collection strategy. Detailed experiments show that: 1) incorporating depth map significantly improves the performance~(we provide result interactive visualization code in supplementary material); 2) \emph{\name} demonstrates robustness to ambient noise interference and RGB-D measurement inaccuracy, showing its potential to be applied in real-scenarios. 
\vspace{-3mm}
\section{Related Work}
\vspace{-2mm}
\begin{figure*}[t]
    \centering
    \includegraphics[width=0.99\linewidth]{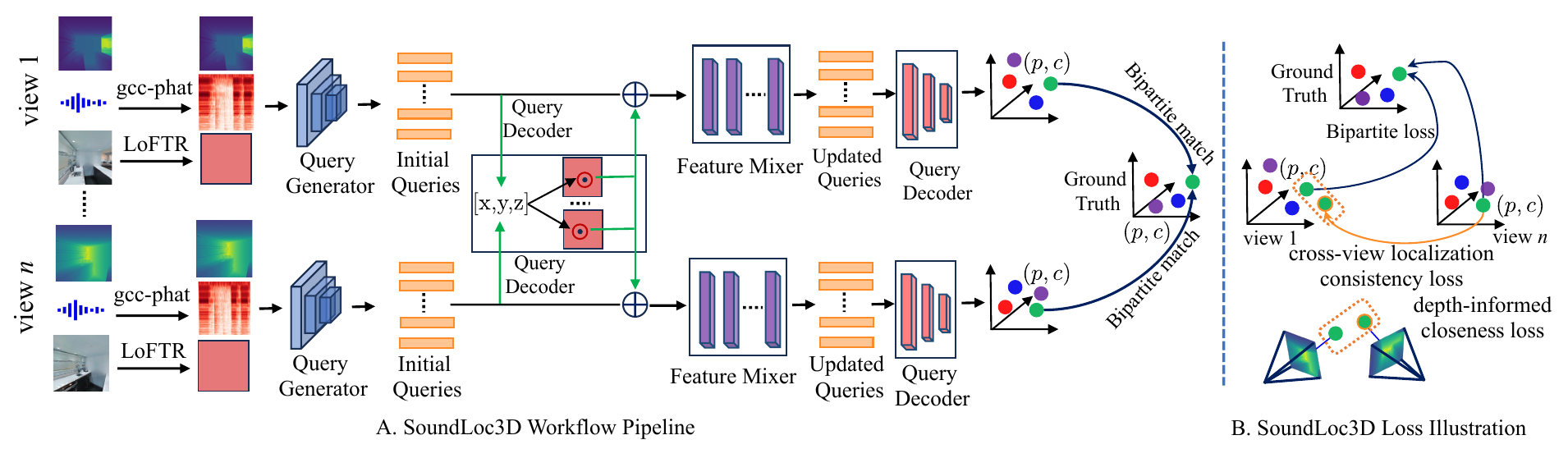}
    \caption{\textbf{SoundLoc3D Pipeline}. The RGB image is first pre-processed by a feature matching aware pre-trained model to get an embedding~(LoFTR), Mic-Array signal feature is extracted by stacking Log-Mel scale TF and GCC-Phat features. The query generator $\mathbfcal{G}$ is applied to get the initial queries, which are further fed to query decoder $\mathbfcal{D}$ to aggregate crossview RGB image informed sound source cues. The queries after aggregation is further optimized by Feature Mixer network~$\mathbfcal{M}$. During training, these queries are matched with ground truth through bipartite matching and the loss considers the discrepancy between prediction and ground truth, depth map informed closeness, and multiview detection consistency. During inference, these optimized queries are simply decoded into sound sources.}
    \label{fig:algo_pipeline}
    \vspace{-5mm}
\end{figure*}

\textbf{Sound Source Detection.} There are several prior works focusing on 3D sound source detection purely from microphone array signals~\cite{l3da_challenge,sounddoa,sounddet,ein_v2,ein,seldnet}. They either detect 3D sound source direction of arrival~(DoA)~\cite{sounddet,ein_v2,seldnet,sounddoa} or spatial position $[x,y,z]$~\cite{l3da_challenge}. In their setting, they assume the microphone receivers are stationary while the sound source can freely move around. This is different from our setting where we instead assume the microphones are movable and the number of sound sources may vary. The recent work Sound3DVDet~\cite{He_2024_WACV} is the most similar work to ours, it involves multiview RGB to assist the localization.

\textbf{Multiview based Object Detection.} Many existing works on multiview based object detection share the core concept proposed in DETR~\cite{DETR}, where object proposals are learned with Transformer in the 2D image space. Adhering to Transformer architecture, DETR3D~\cite{detr3d} expands the domain to 3D with multiview input for learning sparse object queries. They detect by either detecting in polar coordinates~\cite{chen2022polar} or integrating 2D features into 3D domain~\cite{liu2022petr,liu2022petrv2}.

\textbf{Audio-Visual Multimodal Learning.} Audio-visual multimodal learning has received lots of attention in recent years~\cite{gabbay2018seeing,ephrat2018looking,afouras2018conversation,lu2018listen,morrone2019face}, audio-visual dereverberation~\cite{chen2023av_dereverb}, localization and navigation~\cite{gao2018learning, senocak2018learning,tian2018audio,pu2017audio}, mono-to-binaural audio generation. Similar to ours, crossmodal RGB image and depth are incorporated for tasks such as dereverberation~\cite{chen2023av_dereverb} and mono-to-binaural audio generation~\cite{parida2022beyond}.

\textbf{Image Feature Matching.} Finding correspondence between images has been a fundamental topic in computer vision. The research can be divided into detector-based and detector-free methods. Detector-based methods make use of detector to find key-points~\cite{tyszkiewicz2020disk, sarlin2020superglue, pei2022Learning, shen2023detector,bagad2023c, chen2022guide, bellavia2022sift}. Detector-free-based methods find denser correspondences~\cite{rocco2020efficient, li2020dual, liu2020extremely, truong2020glu,sun2021loftr}. We utilize the image matching features to provide sound source localization cue.
\vspace{-3mm}
\section{SoundLoc3D Framework Introduction}
\vspace{-1mm}
\subsection{Problem Definition}
\vspace{-1mm}

We assume $M$ sound sources $\mathcal{S} = \{(p_m, c_m)\}_{m=1}^M$ arbitrarily lie on some physical objects' surfaces in an enclosed room environment, continuously emitting sound waveforms. The physical objects are commonly seen indoor objects such as chair, wall, and door. Each sound source is associated with a 3D spatial position $p_m\in \mathbb{R}^3$ and a semantic class label $c_m \in C$~($C = \{c_1, c_2, \cdots, c_k\}$, $k$ is class number). The sound sources are invisible and mutually independent, thus may lie on the surface of any object in the scene, emitting a sound waveform of any sound class. The task is to jointly localize each sound source's 3D spatial position and predict its semantic label.

\noindent \textbf{RGB-D Acoustic-Camera.} We base our study on the RGB acoustic camera rig proposed in~\cite{He_2024_WACV}, and further equip it with a depth sensor, thereby advancing it to collect depth maps alongside the RGB image.  Specifically, this data acquisition rig consists of a centered pinhole RGB-D camera and four microphones arranged co-planarly at the four corners with 10~cm spacing distance~(see Fig.~\ref{fig:teasing_img1} A). The RGB-D camera and the four microphones are pre-calibrated and synchronized so that we are able to use the rig to record the RGB-D image and Mic-Array signal simultaneously from any viewpoint with known camera poses. In this work, we assume that our method has a coarse estimation of the spatial location of the sound sources~(\eg, the gas pipes run along the walls in a kitchen and the \emph{leak sound} thus comes from the kitchen wall). We use the RGB-D acoustic-camera to record the acoustic scene~\footnote{An acoustic scene indicates a localized area containing the physical object and associated sound sources to be localized.} from $N$ nearby views, denoted $\{(\mathcal{A}_i,I_i, D_i) | T_i\}_{i=1}^N$. For the $i$-th view, Mic-Array signal is denoted by $\mathcal{A}_i = [a_{i, 1}, a_{i, 2}, a_{i, 3}, a_{i, 4}]$, RGB image by $I_i$, depth map by $D_i$, camera pose by $T_i$. Our goal is to design a framework $\mathbf{\Omega}$ to jointly localize and classify 3D sound sources from multiview RGB-D and Mic-Array recordings,

\begin{equation}
     \mathcal{S} \leftarrow \mathbf{\Omega}(\{(\mathcal{A}_i,I_i,D_i) | T_i\}_{i=1}^N).
\end{equation}

To reflect the real scenario, the framework $\mathbf{\Omega}$ must consider several factors: \textbf{c1}, capable of addressing audio-visual weak-correlation, which implies the presence of a 3D sound source is independent on physical object's category; \textbf{c2}, accommodate arbitrary number of 3D sound sources locations. and \textbf{c3}, robust to multiview RGB-D measurement inaccuracies and ambient noise interference. We show how \emph{\name} is designed to be compliant with all these factors.

\begin{figure*}[t]
    \centering
    \includegraphics[width=0.99\linewidth]{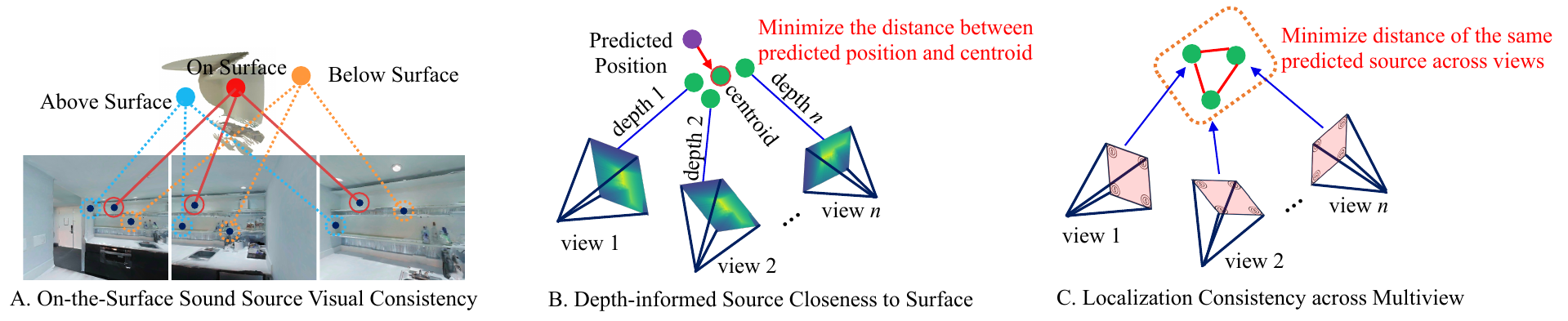}
    \vspace{-1.5mm}
    \caption{\textbf{Sound Source Cue from Multiview RGB-D images and Crossview Consistency:} A. While only ``on the surface'' sound source's projections onto multiview RGB images are guaranteed to be visually similar, either above or below the surface sound sources are much less likely to be visually similar. B. The closer of predicted sound source to the object surface, the smaller of its distance to multiview depth maps informed source position~(centroid). C. The same sound source predicted by each single view should be close enough across views.}
    \label{fig:extracue_vis}
    \vspace{-4mm}
\end{figure*}

Following~\cite{He_2024_WACV}, we formulate \emph{\name} as a \textit{set prediction} problem, where each element in the set corresponds to a potential sound source with unique and stationary spatial position and semantic label. It is worth noting that treating it as \textit{set prediction} problem
can be easily scaled up to handle arbitrary 3D sound sources~(\textbf{c2}). It also avoids us from performing various time-consuming post-processing~(\eg, non-maximum suppression~(NMS)). Following the terminology in recent works~\cite{detr3d,DETR}, we call each element in the \textit{set} as a sound source \textit{query}. The initial queries in the set are learned from each single view Mic-Array signal independently~(\textbf{c1}), which are subsequently optimized by actively incorporating cross-modal sound source cues informed by multiview RGB-D images~(see Fig.~\ref{fig:extracue_vis}). We incorporate three cross-modal sound source cues: 1) visual appearance consistency on the location of the source from multiview RGB images, 2) proximity of the source to an object surface from the multiview depth maps, and 3) cross-view estimation consistency. Specifically, \emph{\name} consists of three main learnable components $\mathbf{\Omega}=(\mathbfcal{G}, \mathbfcal{M}, \mathbfcal{D})$: query generator~$\mathbfcal{G}$ that is responsible for sound source query generation, a feature mixer~$\mathbfcal{M}$ that efficiently integrates multiview cross-modal RGB-D informed sound source cues and a query decoder $\mathbfcal{D}$ that decodes a query into its spatial position and semantic label. The overall pipeline is shown in Fig.~\ref{fig:algo_pipeline}.
\vspace{-2mm}
\subsection{Initial Query Learning from each Single-view Mic-Array Signal}
\label{sec:micarray_querylearn}

The acoustic-camera records four-channel Mic-Array signal from each single view. The four-channel sound waveforms provide enough cues to estimate a sound source's 3D spatial position and semantic label. The time-frequency representation obtained from short time Fourier transform~(STFT) reveals the semantic label and inter-channel phase difference encodes its spatial position. Given one Mic-Array signal $\mathcal{A}_i = [a_{i,1}, a_{i,2}, a_{i,3}, a_{i,4}]$, we follow the practice in~\cite{seld_dcase19,seldnet, resnet18_seld,ein_v2,l3da_challenge} to jointly encode the time-frequency representation in log-mel scale for each single channel waveform, as well as the generalized cross-correlation phase transform~(GCC-Phat~\cite{gcc_phat}) between each pair of channels. The GCC-Phat feature is widely used to encode inter-channel phase difference~\cite{ein_v2,seldnet,time_domain_gcc,ein} as it is relatively insensitive to the ambient noise interference~\cite{gcc_phat}. Given two channel sound waveforms $a_{i, k}$ and $a_{i, l}$ in $\mathcal{A}_i$, the GCC-Phat $f_{{\rm gccphat}, i}^{k,l}$ is,

\vspace{-2mm}
\begin{equation}
    f_{{\rm gccphat}, i}^{k,l} = {\rm ifft}\left(\frac{F(a_{i, k})\cdot F^*(a_{i,l})}{|F(a_{i, k})|\cdot|F^*(a_{i, l})|}\right),\ \  k\neq l,
\end{equation}
\noindent where ${\rm ifft}(\cdot)$ indicates inverse short time Fourier transform, $F(\cdot)$ represents short-time Fourier transform~(afterwards transformed to Log-mel scale), and $F^*$ indicates the complex conjugate, for $k, l \in \{1,2,3,4\}$. Given the four-channel sound waveforms from a single view, we can extract 10 2D feature maps by stacking 4 STFT representations in Log-mel scale and 6 GCC-Phat feature maps~(${4 \choose 2} = 6$) together, $f_{{\rm mic},i} \in \mathbb{R}^{10\times H_1 \times W_1}$ (in our case, $H_1 = W_1 = 256$).

The source query generator $\mathbfcal{G}$ then takes as input the 10-channel feature map $f_{\rm mic}$ to learn the vision-agnostic initial queries $\mathcal{Q}_{\rm init} \in \mathbb{R}^{q\times d}$~(in our case, $q=16$, $d=256$). It is achieved by applying a sequence of 2D convolutions to consecutively reduce the feature map spatial resolution while increasing channel dimension size~(2D convolution with \textit{stride} of 2 to halve the spatial size),
\begin{equation}
\mathcal{Q}_{{\rm init}, i} = \mathbfcal{G}(f_{{\rm mic},i}), \forall i = 1, \cdots, N,
\label{eqn:micgen}
\end{equation}

\noindent where $\mathcal{Q}_{{\rm init}, i}$ indicates the $i$-th view initial source queries, each of which corresponds to a potential  sound source with specific spatial position expressed in its own camera coordinate system~(the $i$-th view camera's coordinate system) and semantic label. We use the query decoder $\mathbfcal{D}$ to decode each query representation into its spatial position and class label,
\begin{equation}
    (P_{{\rm init}, i}, C_{{\rm init}, i}) = \mathbfcal{D}(\mathcal{Q}_{{\rm init}, i}), \forall i = 1, \cdots, N.
    \label{eqn:init_decoder}
\end{equation}
\noindent Rather than directly predicting sources from the initial queries~(Eqn.~\ref{eqn:init_decoder}). We further optimize the queries by incorporating sound source cues from multiview RGB-D images.

\vspace{-1mm}
\subsection{Cross-View Source 3D Position Transform}

To incorporate multiview cross-modal source cues, we transform the decoded position in Eqn.~\ref{eqn:init_decoder} that is expressed in its own camera coordinate system, to the coordinate system of another camera~(\eg, $j$-th view, $i\neq j$). This is achieved by applying rigid transformation with known extrinsic projection matrix $T_{j \leftarrow i}\in \mathbb{R}^{4\times4}$ that translates and rotates from the $i$-th view $P_{{\rm init},i}$ to $j$-th view coordinate system,

\begin{equation}
    P_{{\rm init}, j\leftarrow i}^{T} =  T_{j\leftarrow i} \cdot P_{{\rm init}, i}^{T},\ \ j\neq i,
\label{eqn:coord_trans}
\end{equation}

\noindent where $P_{{\rm init}, i}^{T}\in \mathbb{R}^{4}$ is a transposed $P_{{\rm init}, i}$ in homogeneous coordinates with weight 1. After cross-view 3D position transformation, we can project the decoded sound source's 3D position, expressed in the novel view, to its corresponding RGB image and depth map plane to obtain its 2D projection $[u_x, u_y]_{j\leftarrow i}$ under the perspective projection,

\begin{equation}
    [u_x,u_y]_{j\leftarrow i} = K_j \cdot P_{{\rm init}, j\leftarrow i}^{T}
\label{eqn:2Dproj}
\end{equation}

\noindent where $K_{j} \in  \mathbb{R}^{3\times4}$ is the intrinsic matrix for $j$-th view. Please note that $K_{i}=K_{j}$ as $i$-th and $j$-th views are taken from a single camera.  With the obtained cross-view 2D projection $[u_x, u_y]_{j\leftarrow i}$, we are able to extract cross-view RGB-D images informed sound source cues.

\vspace{-1mm}
\subsection{Multiview RGB Informed Sound Source Cue}
\vspace{-1mm}
\label{rgb_loftr}

Due to the audio-visual weak-correlation, we cannot directly detect a 3D sound source from a single RGB image. Multiview RGB images, however, can be exploited to implicitly provide sound source position constraints. Specifically, based on multiview geometry~\cite{detr3d,chen2022graph,liu2022petr,liu2022petrv2}, an ``on-the-surface'' 3D point's projections onto multiview RGB images are matching points that are visually similar, while either ``above-the-surface'' or ``below-the-surface'' 3D point's projections are non-matching points and thus less visually similar~(see Fig.~\ref{fig:extracue_vis} A)~\cite{zhu2023pmatch, xie2023deepmatcher, tyszkiewicz2020disk}. Depending on this constraint, we can encourage the predicted sound source's spatial position to lie on an object's physical surface and further update the query accordingly. After training, the framework tends to predict query producing the source with matching points projects onto multiview RGB images.

We model crossview visual consistency in feature space. We adopt the pre-trained image matching network LoFTR~\cite{sun2021loftr} to directly provide feature embeddings for each sound source projections onto the multiview RGB images. The LoFTR model is specifically trained for image matching in a coarse-to-fine manner. We use only its coarse-level representation~(of size $\mathbb{R}^{256\times 64 \times 64}$). Given a projection pixel point $[u_x, u_y]$, we use bilinear interpolation to obtain its visual appearance embedding $f_{[u_x, u_y]}$,

\begin{equation}
    f_{[u_x, u_y]} = \phi_{\rm {bilinear}}({\rm LoFTR}(I))_{[u_x, u_y]},
    \label{eqn:rgb_bilinear}
\end{equation}

\noindent where ${\rm LoFTR}(I)$ indicates extracting the LoFTR coarse-level image matching feature representation from the RGB image $I$. For those invalid projections that are off the image plane, we fill its feature with zeros. Finally, for the initial queries from any single view Mic-Array signal, we aggregate its multiview RGB images-informed sound source cues as:

\vspace{-2mm}
\begin{equation}
    \mathcal{Q}_{{\rm init}, i} \leftarrow \mathcal{Q}_{{\rm init}, i} + \frac{1}{N}\sum_{j=0}^{N} f_{[u_x, u_y]_{j\leftarrow i}}; i, j = 1 \cdots, N,
\label{eqn:query_update}
\end{equation}
\vspace{-4mm}

\noindent where $f_{[u_x, u_y]_{j\leftarrow i}}$ indicates the feature extracted in the $j$-th view for the query in the $i$-th view. The updated queries are further fed to $\mathbfcal{M}$ for further optimization.

\vspace{-1mm}
\subsection{Feature Mixer for Query Optimization}
\vspace{-1mm}

The feature mixer $\mathbfcal{M}$ is a Transformer encoder network~\cite{att_all_need}. The updated queries in Eqn.~\ref{eqn:query_update} are flattened into tokens and passed through $\mathbfcal{M}$ for further optimization. The motivation for designing $\mathbfcal{M}$ as a Transformer encoder is two-fold: 1) The $\mathbfcal{G}$ learned queries are order-less and thus naturally fits for Transformer-based network architecture as all tokens are kept order-less during learning; 2) The updated queries carrying multiview cross-modal sound source cues can easily be further optimized by inter-query interaction and per-query learning:

\vspace{-1mm}
\begin{equation}
        \mathcal{Q}_{{\rm update}, i} = \mathbfcal{M}(\mathcal{Q}_{{\rm init}, i} ), \ \ i = 1, \cdots, N
\label{eqn:feat_mixer}
\end{equation}
Given the updated queries $\mathcal{Q}_{{\rm update}}$, we can again apply the query decoder $\mathbfcal{D}$ to decode each individual query into its corresponding spatial position and semantic class label,

\vspace{-1mm}
\begin{equation}
    (P_{{\rm update}, i}, C_{{\rm update}, i)} = \mathbfcal{D}(\mathcal{Q}_{{\rm init}, i}), \ \ i = 1, \cdots, N
    \label{eqn:update_decoder}
\end{equation}

For each decoded query in Eqn.~\ref{eqn:update_decoder}, we apply bipartite matching~\cite{Kuhn1955} to associate it with a ground truth sound source. Since the number of predicted queries is usually larger than the ground truth number, we explicitly append \texttt{non-source} categories $\varnothing$ to the ground truth so as to match the query number. After bipartite matching, we compute $\ell_1$ loss for spatial position regression and cross-entropy loss for semantic label classification.

\vspace{-2mm}
\begin{equation}
    \mathcal{L}_{{\rm bm},i} = \frac{1}{N}\sum_{i=1}^N|P_{{\rm pred}, i} - P_{{\rm gt}, i}| + {\rm CE}(C_{{\rm pred},i}, C_{{\rm gt},i}),
\label{eqn:bm_loss}
\end{equation}

\noindent where $|\cdot|$ and ${\rm CE}(\cdot)$ indicate the $\ell_1$ and the cross-entropy loss, respectively. $P_{{\rm gt}, i}$ and $C_{{\rm gt},i}$ are the matched ground truth sound source spatial position and class label, respectively.

\vspace{-1mm}
\subsection{Multiview Depth Informed Sound Source Cue}
\vspace{-1mm}

Comparing with RGB images, depth map provides more straightforward and direct sound source spatial position cues as each depth map directly indicates the spatial locations of the surface of physical objects. Our key insight is that the query decoder $\mathbfcal{D}$ decoded sound source position's proximity to an object surface can be directly informed by multiview depth maps: projecting the decoded sound source to one view depth map plane to get the projection point, reversing back along the projection ray of the corresponding depth value distance from the projection point naturally gets that depth map informed sound source position. When projecting this query decoded sound source to multiview depth maps, we can accordingly get multiple such depth map informed sound source positions. 

The closer the decoded source position of a query is to the physical object's surface, the more spatially aligned its projection points will be on the multiview depth maps after reprojecting them into 3D space. Conversely, if the source position is farther away, the projection points will be more spatially distant. We thus introduce depth map informed spatial closeness loss to encourage the query-decoded sound source to ``march towards'' physical object's surface~(see Fig.~\ref{fig:extracue_vis} B). In this work, we measure the discrepancy between query-decoded spatial position and the centroid of the multiview depth maps re-projected positions. A loss is incurred if the discrepancy exceeds a pre-defined distance threshold $\sigma$~(in our case $\sigma = 0.3~m$),
\vspace{-1mm}
\begin{equation}
        \mathcal{L}_{{\rm depth}, i} = \max \{||P_{{\rm pred}, i}^{+} - P_{{\rm centroid}, i}||_2 - \sigma, 0\}
        \label{eqn:depth_loss}
\end{equation}
\noindent where $P_{{\rm centroid}, i}$ is the centroid of multiview depth-information sound source positions. $P_{{\rm pred}, i}^{+}$ indicates the decoded queries matched with meaningful ground truth~(not the appended \texttt{no-source}). $||\cdot||_2$ is the $\ell_2$ distance. It is worth noting that the loss is only incurred iff the discrepancy exceeds the distance threshold $\sigma$, which reconciles the depth recording inaccuracies. The multiview depth maps thus update the query to lie closer to the physical surface by directly affecting the loss value.

\vspace{-2mm}
\subsection{Crossview Estimation Consistency}
\vspace{-1mm}
\label{sec:cv_cons_loss}
As shown in Eqn.~\ref{eqn:update_decoder}, we predict the same sound sources from each single view separately. During training stage, each ground truth sound source is matched with one query from each single view and compute the loss~(Eqn.~\ref{eqn:bm_loss}) between the ground truth and each matched query separately. This loss fails to take the detection consistency across multiviews into account, it merely takes the difference between ground truth and each single view detection separately. To enforce the crossview estimation consistency, we explicitly incorporate crossview estimation consistency loss $\mathcal{L}_{{\rm crossview}}$ to force the the same sound source estimated from multiviews to be as spatially close as possible~(see Fig.~\ref{fig:extracue_vis} C),

\vspace{-3mm}
\begin{equation}
    \mathcal{L}_{{\rm crossview}} = \frac{1}{C}\sum_{i=1}^{N}\sum_{j=i}^{N} ||P^{+}_{{\rm pred},i} - T_{i\leftarrow j}\cdot P^{+}_{pred, j}||_{2}, \ \ \ i\neq j
\end{equation}

\noindent where $P_{{\rm pred}, i}^{+}$ and $P_{{\rm pred}, j}^{+}$ indicate the decoded sound source spatial position in Eqn.~\ref{eqn:update_decoder} for the same ground truth sound source from different views. $C$ indicates the view pair combination number, $C={N \choose 2}$.

\vspace{-1mm}
\subsection{Training and Inference}
\vspace{-1mm}

The overall pipeline of our proposed \emph{\name} is shown in Fig.~\ref{fig:algo_pipeline}. Given the collected RGB-D and Mic-Array recordings from multiview, we first extract Log-mel scaled STFT and GCC-Phat feature from four-channel Mic-Array signal~(Sec.~\ref{sec:micarray_querylearn}) and LoFTR~\cite{sun2021loftr} pre-trained image matching model processed RGB image feature embedding~(see Sec.~\ref{rgb_loftr}). Afterwards, the learnable query generator $\mathbfcal{G}$ is applied to learn initial queries representation from each single view Mic-Array signal. Each individual query from one view actively aggregates multiview RGB images informed sound sound cues from all mutliviews by first decoding the query into spatial position and further projecting it to the RGB image plane to collect the corresponding feature. This is achieved by first passing the query feature to the Query Decoder $\mathbfcal{D}$ to get its decoded spatial position and then projecting the spatial position to the corresponding RGB image plane with relevant camera poses. After merging RGB images informed sound source cues in feature space, the initial queries are fed to feature mixer $\mathbfcal{M}$ for further optimization. The optimized queries are again fed to Query Decoder $\mathbfcal{D}$ to be decoded into corresponding spatial position and semantic label. Finally, bipartite matching is applied to match with the corresponding ground truth sound source with one decoded sound source from each view. The overall loss consists of the sum of all of the four aforementioned losses,

\vspace{-4mm}
\begin{equation}
    \mathcal{L} = \lambda_1\cdot \mathcal{L}_{{\rm bm}} + \lambda_2\cdot \mathcal{L}_{{\rm depth}} + \lambda_3\cdot\mathcal{L}_{{\rm crossview}},
\label{eqn:final_loss}
\end{equation}

\noindent where $\lambda_1, \lambda_2, \lambda_3$ are the loss weight and they are all set as 1.0. During training, we adopt the deep supervision strategy~\cite{deepexplorer,deeply_supervised_nets} to jointly supervise the initial queries in Eqn.~\ref{eqn:init_decoder} and updated queries in Eqn.~\ref{eqn:update_decoder} with the loss expressed in Eqn.~\ref{eqn:final_loss}~(which means $P_{{\rm pred}}$ is replaced by $P_{{\rm init}}$ and $P_{{\rm update}}$ separately). During test, we get the query predictions from each single frame and evaluate against ground truth separately. Finally, we add the evaluation result from each single view together. We do not explicitly merge the predictions from multiviews because predictions from different views can be different~(\eg, as it is \textit{set prediction}, the predicted sound source number from different views may vary).
\vspace{-1mm}
\section{Experiments}
\vspace{-1mm}

 \begin{table}[t]
    \scriptsize
    \centering
    \caption{Inference time and param. size. Inference time is tested on Intel Core i9-7920X CPU by averaging 100 independent tests.}
    \vspace{-2mm}
    \begin{tabular}{ l|p{0.6cm}<{\centering}p{0.5cm}<{\centering}|l|p{0.6cm}<{\centering}p{0.5cm}<{\centering} }
      \hline
      \scriptsize Method & Inference & Param. & \scriptsize Method & Inference& Param.\\
      \hline
      SoundDet~\cite{sounddet} & 1.25~s & 13~M&
      EIN-v2~\cite{ein_v2} & 2.20~s & 26~M \\
      SoundDoA~\cite{sounddoa} & 2.10~s & 27~M &
      SALSA~\cite{salsa} & 1.77~s & 11.6~M \\
      SALSA-Lite~\cite{salsa_lite} & 1.37~s & 7.9~M &
       SELDNet~\cite{seldnet} & 1.40~s & 0.7~M \\
       \cline{4-6}
      Sound3DVDet~\cite{He_2024_WACV} & 2.77~s & 19.8~M &
      \textbf{\emph{\name}} & \cellcolor{topcolor}\textbf{1.50~s} & \cellcolor{topcolor}\textbf{3.8~M} \\
      \hline
    \end{tabular}
    \label{table:inference_time_mainpaper}
    \vspace{-4mm}
\end{table}

\textbf{Dataset Creation:} We follow Sound3DVDet~\cite{He_2024_WACV} data creation pipeline to create a large-scale synthetic dataset using the SoundSpaces~2.0~\cite{chen22soundspaces2} and Matterport3D scenes~\cite{Matterport3D}. Specifically, we employ five sound source classes: \textit{telephone-ring}, \textit{siren}, \textit{alarm}, \textit{fireplace} and \textit{horn-beeps} and six physical objects: \textit{wall}, \textit{chair}, \textit{table}, \textit{door}, \textit{ceiling}, and \textit{cabinet}. For any given room scene, we first randomly select a set of those six physical objects. 
For each object, we independently place $n$~($1 \leq n \leq 10 $) sound sources on its surface~(by ensuring any two sources are at least $0.3~m$ apart), each of which isotropically emits sound waveform. Such an object and the placed multiple sound sources are called an acoustic scene. The collected acoustic scenes show, 1) large visual variability even for the same object like chair; 2) various sound source number; 3) various sound source class. These variabilities force all methods to enhance their generalization capability.

 \begin{table*}[t]
  \begin{minipage}[t]{0.68\textwidth}
    \scriptsize
    \centering
    \caption{\footnotesize Quantitative results across all six object categories and five sound classes~(left), result on the texture-homogeneous versus texture-discriminative~(right). We do not report standard deviation due to space limit~(all $\leq 0.010$).}
        \begin{tabular}{l|ccc|ccc|ccc}
    \hline
    \multirow{2}{*}{Methods} & \multicolumn{3}{c|}{Overall Result} & \multicolumn{3}{c|}{Texture Homogeneous} & \multicolumn{3}{c}{Texture Discriminative} \\
    \cline{2-10}
     &mAP & mAR & mALE &mAP & mAR & mALE &mAP & mAR & mALE \\
    \hline
    SELDNet~\cite{seldnet} & 0.103 & 0.501 & 0.923 & 0.107  & 0.532 & 0.910 & 0.100 & 0.528 & 0.934\\
    EIN-v2~\cite{ein_v2} & 0.113 & 0.607  & 0.878 & 0.112  & 0.620 & 0.882 & 0.116 & 0.600 & 0.862 \\
    SoundDoA~\cite{sounddoa} & \cellcolor{thirdcolor}0.212  & \cellcolor{thirdcolor}0.762  & 0.800 & 0.225  & 0.773 & 0.821 & 0.224 & 0.748 & 0.819\\
    SALSA~\cite{salsa} & 0.147  & 0.722  & \cellcolor{thirdcolor}0.793 & 0.146 & 0.722 & 0.791 & 0.147 & 0.723 & 0.794 \\
    SALSA-Lite~\cite{salsa_lite} & 0.130  & 0.712  & 0.810 & 0.131 & 0.710 & 0.811 & 0.130 & 0.713 & 0.811 \\
    SoundDet~\cite{sounddet} & 0.120  & 0.674  & 0.823 & 0.121 & 0.675 & 0.822 & 0.120 & 0.674 & 0.823\\
    Sound3DVDet~\cite{He_2024_WACV} & \cellcolor{secondcolor}0.309  & \cellcolor{secondcolor}0.998  & \cellcolor{secondcolor} 0.586 & 0.308 & 0.997 & 0.584 & 0.296 & 0.992 & 0.589\\
    \hline
    SoundLoc3D & \cellcolor{topcolor}\textbf{0.518}  & \cellcolor{topcolor}\textbf{0.999}  & \cellcolor{topcolor}\textbf{0.320} & \textbf{0.517} & \textbf{0.997} & \textbf{0.312} & \textbf{0.519} & \textbf{0.997} & \textbf{0.301} \\
    \hline
    \end{tabular}
    \label{table:overall_rst_texturecomp}
  \end{minipage}%
  \hfill
  \begin{minipage}[t]{0.28\textwidth}
    \tiny
    \centering
    \captionof{table}{\footnotesize Ablation study on view number. Standard deviation $\leq 0.005$.}
    \vspace{-2mm}
    \begin{tabular}{l|ccc}
    \hline
    View Number & mAP~($\uparrow$) & mAR~($\uparrow$) & mALE~($\downarrow$) \\
    \hline
    1 view & 0.412 & 0.870  & 0.520 \\
    2 views & 0.491  & 0.923  & 0.479 \\
    4 views & \cellcolor{thirdcolor}0.516  & \cellcolor{thirdcolor}0.997  & \cellcolor{thirdcolor}0.320 \\
    6 views & \cellcolor{topcolor}0.522  & \cellcolor{topcolor}0.999  & \cellcolor{secondcolor}0.318 \\
    8 views & \cellcolor{secondcolor}0.521  & \cellcolor{topcolor}0.999  & \cellcolor{topcolor}0.309 \\
    \hline
    \end{tabular}
    \label{table:viewnum_test}
    \vspace{-1mm}
    \tiny
    \centering
    \captionof{table}{\footnotesize Ablation study on model architecture variants. Standard deviation $\leq 0.03$.}
    
    \begin{tabular}{l|ccc}
    \hline
    Methods & mAP~($\uparrow$) & mAR~($\uparrow$) & mALE~($\downarrow$) \\
    \hline
    SL3D\_noRGB & \cellcolor{thirdcolor}0.498  & \cellcolor{thirdcolor}0.944  & 0.510 \\
    SL3D\_noDepth & 0.472 & 0.910 & \cellcolor{thirdcolor}0.457 \\
    SL3D\_noCVC & \cellcolor{secondcolor}0.501  & \cellcolor{secondcolor}0.948  & \cellcolor{secondcolor}0.389 \\
    SL3D\_noRGBD &  0.328  & 0.732 &  0.810\\
    \hline
    SoundLoc3D  & \cellcolor{topcolor}\textbf{0.518}  & \cellcolor{topcolor}\textbf{0.999}  & \cellcolor{topcolor}\textbf{0.320}\\
    \hline
    \end{tabular}
    \label{table:overall_rst_ablation}
  \end{minipage}
  \vspace{-2mm}
\end{table*}

During multiview recording, we put the acoustic-camera approximately $3m$ away from the acoustic scene and ensure all sound sources are not visually blocked in any view. Mic-Array sampling frequency is 21k Hz. To test the visual discriminativeness of the RGB images and their impact on the performance, we divide the scenes into two main categories based on sound source projections' onto the multiview images: 1) texture-homogeneous acoustic scene in which the sound source projections lie on homogeneous textured areas~(\eg, textureless wall, table), 2) texture-discriminative acoustic scene where the projections localize on texture discriminative area~(\eg, corner, edge). We created 5,000/1250 acoustic scenes for training and test respectively, after filtering views without any depth map. The acoustic scene variabilities discussed above guarantee the training and test sets exhibit enough visual and acoustic difference.

\noindent\textbf{Evaluation Metrics:} Following~\cite{He_2024_WACV}, we adopt three metrics: mean average precision~(mAP), mean average recall~(mAR), mean average localization error~(mALE). Given the predicted sound source set and ground truth for a particular class, we apply the bipartite matching algorithm~\cite{Kuhn1955} to assign each detected sound source to one ground truth sound source. After the assignment, a detected sound source is a true positive if it is within a distance threshold~(we adopt three distance thresholds: $[0.5~m, 0.8~m, 1.2~m]$) with its assigned ground truth, otherwise it is treated as a false positive. Afterwards, we compute mAP, mAR and mALE~(refer to~\cite{He_2024_WACV}). Higher mAP and mAR and lower mALE indicate better performance.

\noindent\textbf{Comparison Methods:} We compare with 1) six most recent Mic-Array signal based sound source localization and detection baselines: SELDNet~\cite{seldnet}, EIN-v2~\cite{ein_v2} and SoundDoA~\cite{sounddoa}, SoundDet~\cite{sounddet}, SALSA~\cite{salsa}, SALSA-Lite~\cite{salsa_lite}. SELDNet has been used as baseline against various methods, it combines CNN and GRU~\cite{GRU} to infer sound sources; EIN-v2~\cite{ein_v2} and SoundDoA~\cite{sounddoa} are two more recent works; they adopt Transformer~\cite{att_all_need} and permutation invariant training~\cite{pit_permu} to infer the location of sound sources. SALSA~\cite{salsa}, SALSA-Lite~\cite{salsa_lite} propose to extract log-linear spectrograms and normalized principal eigenvector to represent Mic-Array data. 2) one multimodal method Sound3DVDet~\cite{He_2024_WACV}, which is most relevant to our setting. The comparison on parameter size and inference time is given in Table~\ref{table:inference_time_mainpaper}, where we can see \emph{\name} is lightweight and efficient.

\begin{figure*}[t]
    \centering
    \includegraphics[width=0.90\linewidth]{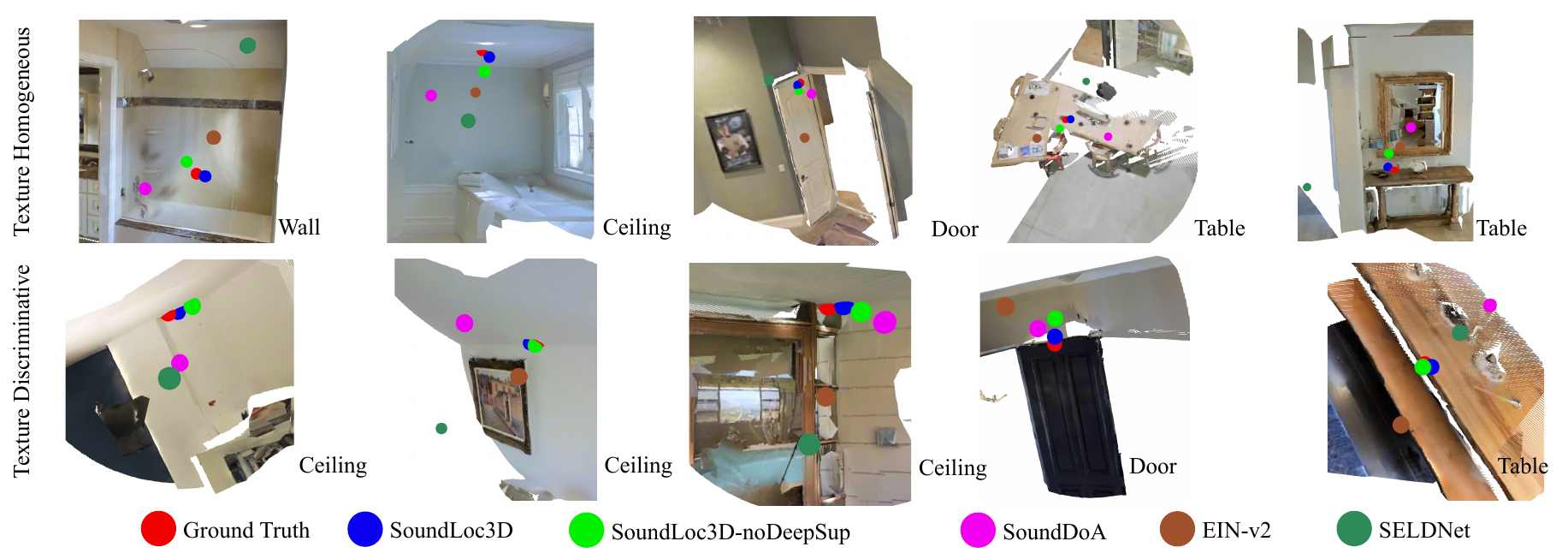}
    \vspace{-2mm}
    \caption{\textbf{Localization Result Visualization}: We visualize the sound source localization result in the 3D visual space by different methods as well as its ground truth position. Zoom in for better visualization. We provide data and visualization code and in Supplementary material.}
    \vspace{-2mm}
    \label{fig:qual_vis}
\end{figure*}

\noindent\textbf{Implementation Details:} Our framework is implemented in PyTorch~\cite{pytorch} and the source code is provided with the supplementary materials. For training the models, we use the AdamW optimizer~\cite{adamW_optimizer} with a learning rate of $0.0001$. Each model is trained for 100 epochs. We train the models three times independently to report the mean and variance for each metric separately. For the Mic-Array signal, the sampling frequency is 21~kHz and we record 1~second data points. The converted time frequency map is of size $256\times 256$ with ${\rm n\_fft} = 511$ and ${\rm hop\_len} = 78$.

\begin{table*}[t]
    \begin{minipage}[t]{0.60\textwidth}
        \scriptsize
        \centering
        \caption{Ablation study on microphone number. CNum indicates the GCC-Phat feature channel number. Standard deviation $\leq 0.02$.}
        \vspace{-3mm}
        \begin{tabular}{l|p{0.5cm}<{\centering}p{0.5cm}<{\centering}p{0.5cm}<{\centering}|p{0.5cm}<{\centering}p{0.5cm}<{\centering}p{0.5cm}<{\centering}|p{0.5cm}<{\centering}p{0.5cm}<{\centering}p{0.5cm}<{\centering}}
        \hline
        \multirow{2}{*}{Method} & \multicolumn{3}{c|}{Mic Num. = 4~(default)} & \multicolumn{3}{c|}{Mic Num. = 6} & \multicolumn{3}{c}{Mic Num. = 8}  \\
        \cline{2-10}
          & mAP & mALE & CNum & mAP & mALE & CNum & mAP & mALE & CNum \\
        \hline
        SoundLoc3D  & 0.520 & 0.313 & 6 & 0.527 & 0.308 & 15 & 0.530 & 0.297 & 28 \\
        \hline
        \end{tabular}
        \label{table:micnum_test} 
    \end{minipage}
    \hfill
    \begin{minipage}[t]{0.37\textwidth}
        \scriptsize
        \centering
        \caption{mAP w.r.t. RGB~(Sound3DVDet) and RGB-D~(SoundLoc3D) measurement inaccuracy.}
        \vspace{-2.5mm}
        \begin{tabular}{ l|p{0.8cm}<{\centering}p{0.7cm}<{\centering}p{0.7cm}<{\centering}p{0.7cm}<{\centering} }
          \hline
          Method & $\delta=0$ & $0.1$ & $0.2$ & $0.3$\\
          \hline
          Sound3DVDet & 0.309 & 0.278 & 0.243 & 0.200 \\
          \hline
          SoundLoc3D  & \cellcolor{topcolor}\textbf{0.516} & \cellcolor{topcolor}\textbf{0.507} & \cellcolor{topcolor}\textbf{0.498} & \cellcolor{topcolor}\textbf{0.480} \\
          \hline
        \end{tabular}
        \label{table:rgbd_measureinaccuracy}
    \end{minipage}
    \hfill
\end{table*}

\vspace{-1mm}
\subsection{Experiment Results}
\vspace{-1mm}

The quantitative results are given in Table~\ref{table:overall_rst_texturecomp}, we can see that \emph{\name} outperforms all the seven comparing methods by a large margin. Comparing with the Mic-Array based best-performing SoundDoA~\cite{sounddoa}, \emph{\name} shows a gain of $0.30$ in mAP, $0.23$ in mAR and $0.48$ in mALE with much smaller network size. Given that most of these methods have larger model sizes~(Table~\ref{table:inference_time_mainpaper}), the efficacy of \emph{\name} is prominent~(even without vision, SL3D\_noRGBD in ablation study). \emph{\name} also outperforms Sound3DVDet~\cite{He_2024_WACV} significantly with much smaller model size.

Further, all methods achieve higher mAR than mAP, suggesting that treating sound source localization and detection as a \textit{set prediction} is capable of estimating all potential sound sources. The overall quantitative performance in terms of texture difference is given in Table~\ref{table:overall_rst_texturecomp} right. We observe that \emph{\name} achieves state-of-the-art performance on both texture homogeneous and discriminative scenes, while the other six Mic-Array only methods show no difference because they do not explicitly leverage vision in their methods. \emph{\name} also outperforms Sound3DVDet~\cite{He_2024_WACV} significantly, showing the potential of depth map in localizing sources. The qualitative comparison is in Fig.~\ref{fig:qual_vis}.

\vspace{-1mm}
\subsection{Ablation Studies}
\vspace{-1mm}

\noindent\textbf{1. Does RGB-D help?} 1) We remove RGB based part in our framework~(Sec.~\ref{rgb_loftr} and only feed the initial queries to feature mixer, \textit{SL3D\_noRGB}; 2) To test the impact of depth maps, we remove the depth map informed loss~(Eqn.~\ref{eqn:depth_loss}) and call this version \textit{SL3D\_noDepth}; 3) We remove both RGB and depth maps to test cross-modal supervision.  From Table~\ref{table:overall_rst_ablation}, we see that removing either RGB images or depth maps results in reduced performance, removing depth maps leads to larger performance drop than RGB images. In \textit{SL3D\_noRGBD}, we observe the largest performance drop. It thus shows, in audio-visual weak-correlation, multiview cross-modal visual information can still be use to significantly assist this task.

\noindent\textbf{2. Does Crossview Estimation Consistency help?} We remove the crossview consistency loss introduced in Sec.~\ref{sec:cv_cons_loss}~(variant \textit{SL3D\_noCVC} in Table~\ref{table:overall_rst_ablation}). The dropped performance confirms the importance of crossview consistency.

\noindent \textbf{3. Microphone Number Impact}. To understand the impact of microphone array number, we collect another three datasets in which all sound sources lie on ``wall'' object surface~(each dataset includes 800 acoustic scenes for training, and 200 for test). The three datasets are identical except that number of microphones used to record the acoustic scene~(we vary the microphone number from 4, 6 to 8). It is worth noting that a minimum of 4 microphones are needed to localize a 3D sound source. In our implementation, all microphones are arranged on a circular plane with a radius 9~cm to the camera center. The results in Table.~\ref{table:micnum_test} show that more microphones can improve the performance, but the performance gain comes with extra computations.

\noindent \textbf{4. View Number Impact}. Note that the above setup uses four views. To assess the influence of the number of views on performance, we further curated a new dataset where we fixed the number of sound sources and classes but changed the number of views in each acoustic scene. Specifically, we involve three sound sources: telephone, siren and alarm. The sources are placed on the ``wall'' and the number of views for each acoustic scene varies from 1 to 8. In total, 1,000 acoustic scenes are generated, with 800/100 split for training and test. Five \textit{Sound3DLoc} models are trained on this dataset using views in $\{1, 2, 4, 6, 8\}$. The results are given in Table~\ref{table:viewnum_test}, and it shows two key points: 1) a conspicuous enhancement in performance as the views increase from 1 to 4, and 2) nearly no performance gain as views continue to increase. This indicates that incorporating multiview recordings is beneficial for the task, but the extent of improvement soon diminishes as more views are incorporated.

\vspace{-2mm}
\subsection{Robustness of the Framework}
\vspace{-2mm}

\begin{table}[t]
    \begin{minipage}{0.20\textwidth}
        \centering
        \scriptsize
        \begin{tabular}{c|cc}
        \hline
        Num & Sound3DVDet & Ours  \\
        \hline
        5 & 0.267 & 0.497 \\
        7 & 0.255 & 0.490 \\
        9 & 0.254 & 0.490 \\
        \hline
        \end{tabular}
        \caption{\scriptsize mAP~($\uparrow$) w.r.t. number of source classes. we run the experiment on 400-train, 100-test dataset by increasing sound class number to 7 and 9~(added bird/engine/turbine/fan sound). We can conclude that increasing source class number leads to negligible performance reduction, showing \emph{\name} is capable of handling multiple sound source classes situation.}
        \label{table:classnum_test}
    \end{minipage}
    \hspace{0.007\textwidth}
    \begin{minipage}{0.28\textwidth}
        \centering
        \includegraphics[width=0.99\textwidth]{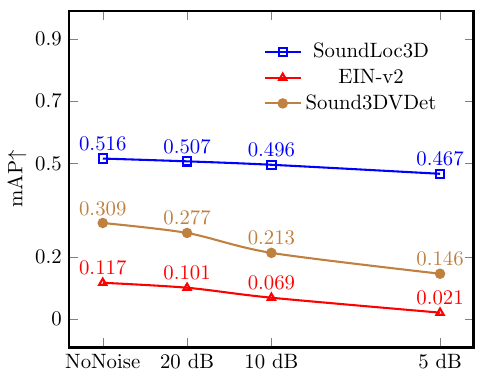}
      \captionof{figure}{\scriptsize \textbf{Ambient noise test}: we add white Gaussian ambient noise measured by SNR in dB.}
      \label{fig:ambient_noise_test}
    \end{minipage}
\vspace{-3mm}
\end{table}

While it is preferable to test our framework on a real-world dataset, collecting such data is both technically and practically challenging. For example, it is difficult to place the sound source on objects' physical surface and it is ``visually invisible''. To this end, we study the robustness in two settings to best imitate the real-world. First, by including additive white Gaussian ambient acoustic noise to the ``wall'' data subset, where  the amount of acoustic noise is measured by signal-to-noise ratio~(SNR, the lower of it, more noise is added). The mAP variations are shown in Fig.~\ref{fig:ambient_noise_test}, where we see both EIN-v2~\cite{ein_v2} and Sound3DVDet~\cite{He_2024_WACV} see a significant drop but \emph{\name} maintains its performance. Second, we add camera pose noise to imitate real-world RGB-D or RGB measurements. Specifically, we add a Gaussian noise $N(0, \delta$)~(mean 0, std $\delta$) to camera rotation parameter~(pitch, roll, yaw) to generate RGB-D or RGB measurement inaccuracy. The result in Table~\ref{table:rgbd_measureinaccuracy} shows the robustness of SoundLoc3D. Third, we add more sound classes: from 5 to 9. The results in Table~\ref{table:classnum_test} show \emph{\name} can handle detection under more sound source classes. In summary, \emph{\name} is capable of localizing and classifying invisible 3D sound sources from multiview RGB-D Mic-Array recordings. It is efficient, scalable and robust to measurement noise and ambient noise interference that are common in real world, demonstrating its potential to be employed in real world.

\textbf{Conclusions and Limitations} we show multiview RGB-D and Mic-Array recordings can be used to estimate invisible sound sources' spatial position and semantic class. Building and experimenting with a real RGB-D acoustic-camera rig is an important future direction.

{\small
\bibliographystyle{ieee_fullname}
\bibliography{main}
}

\newpage
\onecolumn
\appendix
\section*{Appendix}
\setcounter{page}{1}

\section{More Discussion on LoFTR}

\setlength{\intextsep}{10pt}
\begin{wrapfigure}{r}{0.5\textwidth}
    \centering
    \includegraphics[width=0.49\textwidth]{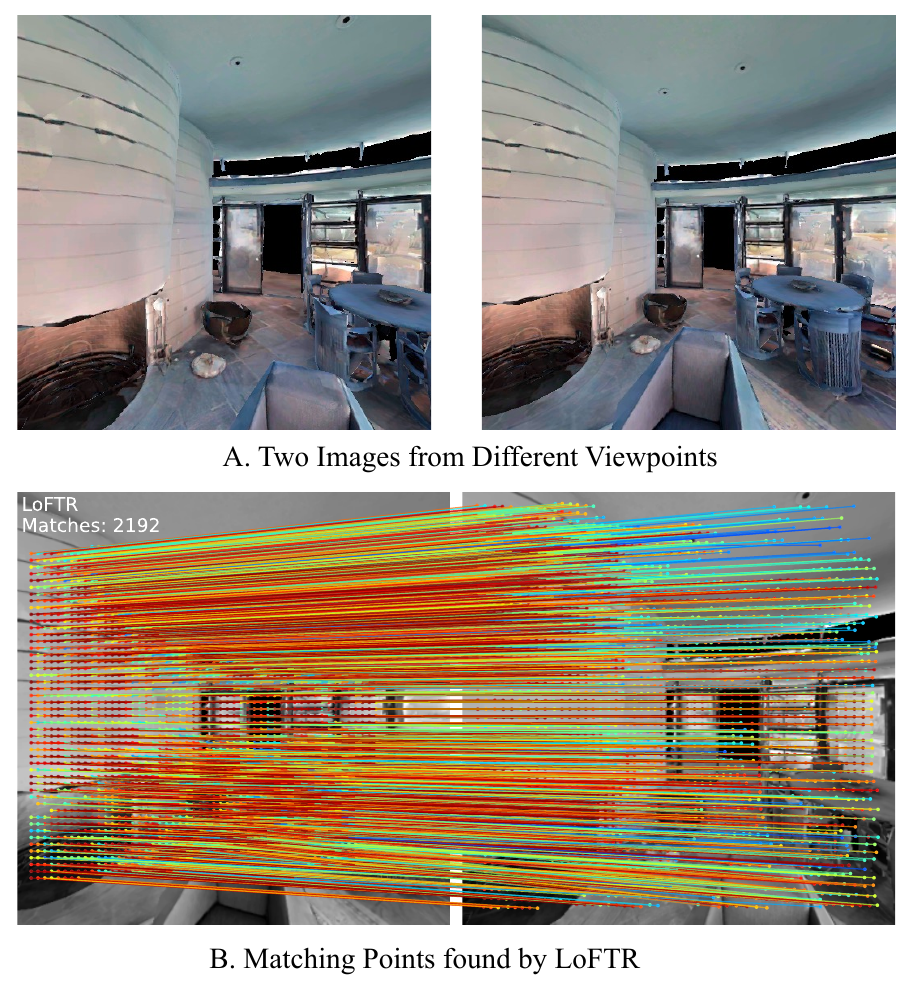}
            \caption{\textbf{Matching points visualization}. \textbf{A.} Two RGB images from different views. They contain large texture homogeneous area. \textbf{B.} LoFTR manages to predict dense matching points even on these texture homogeneous areas.}
    \label{fig:loftr_rep}
\end{wrapfigure}

We adopt LoFTR~\cite{sun2021loftr} to extract RGB image feature, which provides RGB informed sound source ``on-the-surface'' appearance consistency constraint across multiviews. LoFTR~\cite{sun2021loftr} is a feature matching model that provides ``matching point'' across RGB images, it naturally fits for our situation because we depend on such ``matching point'' to infer 3D sound source's spatial localization. Due to its coarse-to-fine learning strategy, LoFTR is capable of retrieving matching points on both texture homogeneous and texture discriminative area. We show such an example in Fig.~\ref{fig:loftr_rep}, from which we can clearly see that dense matching point pairs are generated on the texture homogeneous wall and ceiling area. It thus shows LoFTR~\cite{sun2021loftr} can provide useful sound source clues for 3D sound source position, regardless of the position's visual appearance. In Sec.~\ref{supp:ablation_study}, we show LoFTR RGB image feature extractor generates better performance than ImageNet pre-trained ResNet50~\cite{resnet18} image feature extractor.

\section{Network Architecture}

\label{sec:nn_arch}
\emph{\name} network architecture is given in Table~\ref{table:SoundLoc3D_net}. The trainable parameter size of our network is 3.8~M. It is worth noting that our proposed \textit{SoundLoc3D} framework is scalable. Its model complexity can be easily scaled up by adding, for example, more Feature Mixer layers~(Transformer encoder layer) or increasing the query embedding size.

\begin{table}[h]
	\centering
        \scriptsize
        \caption{\textit{SoundLoc3D} network architecture illustration. In the Query Generator $\mathbfcal{G}$, the 2D convolution kernel size is $3\times 3$ and the stride is 2.}
	\begin{tabular}{c|c|c|c}
    \toprule
    \multicolumn{4}{c}{\textbf{Query Generator $\mathbfcal{G}$}: Input: $[10, 256, 256]$} \\
        \hline
        Layer Name & In-ch. Num. & Out-ch. Num.  & feature size \\
        \hline
        Conv2D & 10 & 32 & $[32, 128, 128]$ \\
        Conv2D & 32 & 64 & $[64, 64, 64]$ \\
        Conv2D & 64 & 128 & $[128, 32, 32]$ \\
        Conv2D & 128 & 256 & $[256, 16, 16]$ \\
        Conv2D & 256 & 512 & $[512, 8, 8]$ \\
        Conv2D & 512 & 256 & $[256, 4, 4]$ \\
        \hline
        \multicolumn{4}{c}{Query Generator Output: $[16, 256]$}\\
        \midrule
        \multicolumn{4}{c}{RGB Informed Feature Aggregation} \\
        \hline
        \multicolumn{4}{c}{LoFT input: $[256, 64, 64]$} \\
        \multicolumn{4}{c}{Aggregation output: $[16, 256]$} \\
        \midrule
        \multicolumn{4}{c}{\textbf{Feature Mixer} $\mathbfcal{M}$} \\
        \hline
        \multicolumn{3}{c|}{Transformer Layer Num} & 1 \\
        \multicolumn{3}{c|}{Token Num} & 16 \\
        \multicolumn{3}{c|}{Head Num} & 4 \\
        \multicolumn{3}{c|}{FFT Dim} & 1024 \\
        \multicolumn{3}{c|}{Output} & $[16, 256]$ \\
        \midrule
        \multicolumn{4}{c}{\textbf{Query Decoder} $\mathbfcal{D}$} \\
        \hline
        \multicolumn{4}{c}{Position Regression Head} \\
        \hline
        Linear + BN + ReLU & 256 & 128 & $[16, 128]$ \\
        Linear & 128 & 3 & $[16, 3]$ \\
        \hline
        \multicolumn{4}{c}{Classification Head} \\
        \hline
        Linear  & 256 & class num & $[16, {\rm class num}]$ \\
        \bottomrule
	\end{tabular}
         \label{table:SoundLoc3D_net}
\end{table}

\section{More Discussion on Dataset Creation}

\begin{figure*}[t]
    \centering
    \includegraphics[width=0.95\linewidth]{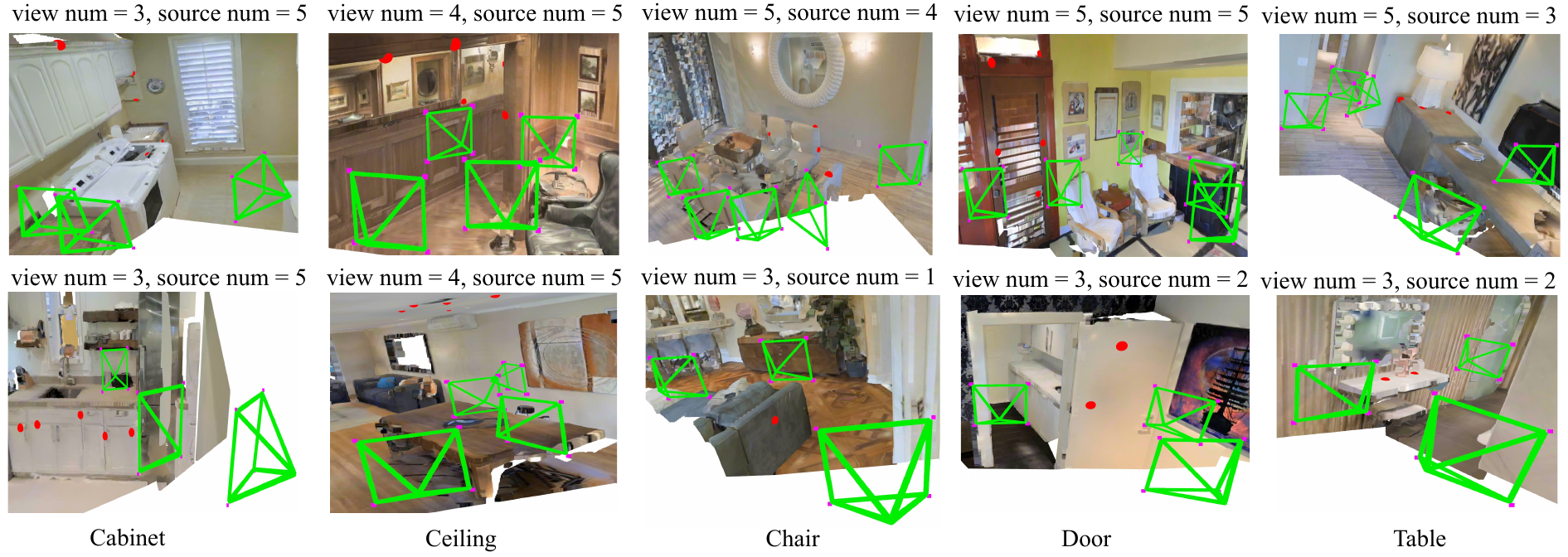}
    \caption{\emph{\name} experiment data visualization: We visualize the sample data we used in our experiment.}
    \label{fig:moredata_vis}
\end{figure*}

In the supplementary material, we provide the statistics of the created large dataset in Table~\ref{table:data_summary} w.r.t. different physical object class. In this table, we can observe that the ``wall'' and ``ceiling'' consist of the largest portion of the dataset, which reflects the real scenario. We further provide some visualizations of the created dataset in Fig.~\ref{fig:moredata_vis}, from which we can have an intuitive understanding of how the dataset look like.

We followed the data creation method introduced in Sound3DVDet~\cite{He_2024_WACV} to create the dataset. We skip the sampled position when no depth map can be collected, so the dataset used by Sound3DVDet~\cite{He_2024_WACV} and this paper is not exactly the same~(the reported Sound3DVDet result in this paper is slightly different from the result reported in the original Sound3DVDet paper). We will release the data creation code and created dataset if this paper is accepted. It is worth noting that, although we placed a 3D sound source on a specific physical object surface in the dataset we have created, the 3D sound sources can freely lie on an arbitrary physical surface. In another word, the sound source placement is independent of physical objects.

\begin{table}[h]
    \centering
    \small
    \caption{Created Multiview Mic-Array and RGBD Dataset Summary w.r.t. each Physical Object Category.}
    \vspace{-3mm}
    \begin{tabular}{l|cccc}
    \toprule
    Object & Texture-homo. & Texture-disc. & Source Num. & View Num.\\
    \midrule
    wall & 975 & 717 & 1-5 & 4\\
    ceiling & 727 & 614 & 1-5 & 4 \\
    table & 464 & 461 & 1-5 & 4\\
    door & 712 & 702 & 1-5 & 4\\
    cabinet & 286 & 292 & 1-5 & 4\\
    chair & 100 & 222 & 1-5 & 4\\
    \midrule
    sum & 3264 & 3008 & / & / \\ 
    \bottomrule
    \end{tabular}
    \label{table:data_summary}
\end{table}

\section{More Details on Train and Test Configuration}

During training, \textit{SoundLoc3D} integrates the predictions~(queries) from each single view by introducing global losses so that \textit{SoundLoc3D} gets optimized by both individual view prediction and crossview prediction consistency. During test, since we adopt \textit{set prediction} strategy to predict sound sources, the sound sources predicted from different views may be different from each other in terms of predicted sound source 3D position class, we treat the \textit{set prediction} from different views separately and compare per-view prediction with ground truth separately to get the final evaluation metric.

\section{More Experiment Result}

We train all models with the same experimental setting presented in the main paper. We train all models three times independently and report the mean value. We do not report the standard deviation because of the space limit in Table~\ref{table:quant_rst_objclass} and Table~\ref{table:quant_rst_ssclass}. All the standard deviations are within $0.02$.

\subsection{More Ablation Study}
\label{supp:ablation_study}

\begin{table}[h]
    \centering
    \small
    \caption{More Ablation Study Quantitative Result.}
    \begin{tabular}{l|ccc}
    \hline
    Methods &\textbf{mAP}~($\uparrow$) & \textbf{mAR}~($\uparrow$) & \textbf{mALE}~($\downarrow$) \\
    \hline
    SL3D\_Res50 & 0.488 $\pm$ 0.002 & 0.932 $\pm$ 0.020 & 0.520 $\pm$ 0.005 \\
    SL3D\_noDeepSup & 0.487 $\pm$ 0.002 & 0.960 $\pm$ 0.001 & 0.391 $\pm$ 0.002\\
    \hline
    Ours Sound3DLoc & \cellcolor{topcolor}\textbf{0.518} $\pm$ 0.010 & \cellcolor{topcolor}\textbf{0.999} $\pm$ 0.001 & \cellcolor{topcolor}\textbf{0.320} $\pm$ 0.001\\
    \hline
    \end{tabular}
    \label{table:supp_ablation}
\end{table}

In the main paper, we reported three ablation studies. We further report another two ablation studies in Table~\ref{table:supp_ablation} to validate the efficiency of SoundLoc3D.

\textbf{1. LoFTR vs ResNet} LoFTR~\cite{sun2021loftr} is better suited to our problem setup  as it uses the projections of sound source locations with visual consistency. We test the performance of replacing LoFTR with widely used ImageNet~\cite{deng2009imagenet} pre-trained ResNet50~\cite{resnet18} as image feature extractor. This variant, we call, \textit{SL3D\_Res50} leads to performance drop as well, which indirectly shows visual consistency is a vital cue for sound source localization.

\textbf{2. Without Deep Supervision.} In \textit{Sound3DLoc}, we jointly train both the initial queries and updated queries. We ablate the performance without deep supervision. To this end, we remove the loss~(Eqn.~(14) in the main paper) added to the initial queries~(\textit{SL3D\_noDeepSup}). From Table~\ref{table:supp_ablation}, we can see that removing deep supervision strategy leads to performance drop.

In summary, from all ablation studies, we can validate the necessity and importance of each component of our \textit{SoundLoc3D} framework design.

\subsection{Quantitative Result w.r.t Sound Source Class}

The quantitative result w.r.t. sound source class is given in Table~\ref{table:quant_rst_ssclass}. From this table, we can observe that 1) \textit{SoundLoc3D} stays as the best-performing method among all comparing methods and all \textit{SoundLoc3D} variants used in ablation studies. 2) As the training dataset size increases~(so the acoustic scenes' visual variation increases accordingly), the three comparing methods have observed performance drop while our proposed \textit{SoundLoc3D} maintains nearly the same performance. It thus shows our proposed \textit{SoundLoc3D} can better handle visual variation challenge. 

\subsection{Quantitative Result w.r.t Physical Object Class}

The detailed quantitative result of our method w.r.t. physical object class is given in Table~\ref{table:quant_rst_objclass}. We can observe from this table that 1) our proposed \textit{SoundLoc3D} outperforms all other \textit{SoundLoc3D} variants across all physical object classes, in terms of both mAP, mAR and mALE metrics. 2) \textit{SoundLoc3D} and its variants achieve better performance on surface flat objects~(such as Table, Ceiling and Wall) than on surface uneven objects~(Chair and Cabinet and Door). It thus shows that localizing and classifying 3D sound sources on cluttered and uneven surface is a challenging task that requires more future work.

\subsection{Comparing Methods with RGBD Image Input}
All the 6 comparing methods~SELDNet~\cite{seldnet}, EIN-v2~\cite{ein_v2}, SoundDoA~\cite{sounddoa}, SoundDet~\cite{sounddet}, SALSA~\cite{salsa}, SALSA-Lite~\cite{salsa_lite} are just based on Mic-Array signal input. One question that naturally arises is that what if feeding the RGBD images to these Mic-Array based methods. To this end, we further run two kinds of extra experiments:

First, we simply combine the Mic-Array signal feature map~($10\times 256 \times 256$) with its corresponding RGBD image~($4\times 256 \times 256$) for each single view. We then obtain a 13-channel 2D Mic-Array and RGBD feature map and feed its neural network to localize and classify sound sources. It helps to test if directly concatenating RGBD can improve Mic-Array based methods' performance. We run such test on SELDNet~\cite{seldnet}, EIN-v2~\cite{ein_v2}, SALSA~\cite{salsa}, SALSA-Lite~\cite{salsa_lite} and Sound3DVDet~\cite{He_2024_WACV}, we do not include SoundDoA~\cite{sounddoa} and SoundDet~\cite{sounddet} because SoundDoA~\cite{sounddoa} and SoundDet~\cite{sounddet} do not directly generate fixed size Mic-Array based 2D feature~(they propose learnable filter bank to directly learn from sound raw waveform). The quantitative result is given in Table~\ref{table:supp:add_rgbd}, from which we can see that simply concatenating RGBD images to Mic-Array feature leads to reduced performance for all comparing methods. It thus shows single view RGBD image does not present explicit 3D sound source localization clue.

\begin{table}[t]
    \centering
    \small
    \caption{Quantitative results of comparing methods w/o single view RGBD input.}
    \begin{tabular}{l|ccc}
    \hline
    Methods &\textbf{mAP}~($\uparrow$) & \textbf{mAR}~($\uparrow$) & \textbf{mALE}~($\downarrow$) \\
    \hline
    SELDNet~\cite{seldnet} & 0.103 $\pm$ 0.002 & 0.501 $\pm$ 0.001 & 0.923 $\pm$ 0.001\\
    SELDNet~\cite{seldnet} + RGBD & 0.093 $\pm$ 0.001 & 0.489 $\pm$ 0.002 & 0.943 $\pm$ 0.001\\
        \hline
    EIN-v2~\cite{ein_v2} & 0.113 $\pm$ 0.002 & 0.607 $\pm$ 0.001 & 0.878 $\pm$ 0.001\\
    EIN-v2~\cite{ein_v2} + RGBD & 0.101 $\pm$ 0.001 & 0.591 $\pm$ 0.001 & 0.899 $\pm$ 0.001\\
    \hline
    SALSA~\cite{salsa} & 0.147 $\pm$ 0.002 & 0.722 $\pm$ 0.002 & 0.793 $\pm$ 0.003\\
    SALSA~\cite{salsa} + RGBD & 0.133 $\pm$ 0.002 & 0.701 $\pm$ 0.001 & 0.813 $\pm$ 0.002\\
    \hline
    SALSA-Lite~\cite{salsa_lite} & 0.130 $\pm$ 0.010 & 0.712 $\pm$ 0.003 & 0.810 $\pm$ 0.002\\
    SALSA-Lite~\cite{salsa_lite} + RGBD & 0.107 $\pm$ 0.006 & 0.697 $\pm$ 0.002 & 0.831 $\pm$ 0.001\\
    \hline
    Sound3DVDet~\cite{He_2024_WACV} & 0.309 $\pm$ 0.010 & 0.998 $\pm$ 0.007 & 0.586 $\pm$ 0.009\\
    Sound3DVDet~\cite{He_2024_WACV} + RGBD & 0.278 $\pm$ 0.009 & 0.892 $\pm$ 0.002 & 0.687 $\pm$ 0.007\\
    \hline
    \emph{\name} & \textbf{0.518} $\pm$ 0.010 & \textbf{0.999} $\pm$ 0.001 & \textbf{0.320} $\pm$ 0.001\\
    \hline
    \end{tabular}
    \label{table:supp:add_rgbd}
\end{table}

Second, we further follow \textit{SoundLoc3D} pipeline to add multiview RGB-informed sound source position visual appearance constraint to the comparing methods. Specifically, we replace \textit{SoundLoc3D}'s Query Generator $\mathbfcal{G}$ with the comparing Mic-Array based methods, and keep the remaining \textit{SoundLoc3D} component the same~(including the Feature Mixer~$\mathbfcal{M}$ and Query Decoder $\mathbfcal{D}$). Such setting helps test the feasibility of our multiview RGBD feature aggregation scheme.  The quantitative result is given in Table~\ref{table:supp:add_mvrgbd}, from which we can see that involving multiview RGB-D informed 3D sound source clue significantly improves their corresponding performance~(in terms of both mAP, mAR and mALE evaluation metric). It thus shows aggregating cross-modal vision-informed clue for sound source localization and classification can dramatically improve the performance, even though the sound source exhibits no visual entity.

\begin{table}[h]
    \centering
    \small
    \caption{Quantitative results of comparing methods w/o multiview RGBD-Informed Sound Source Clue Aggregation.}
    \begin{tabular}{l|ccc}
    \hline
    Methods &\textbf{mAP}~($\uparrow$) & \textbf{mAR}~($\uparrow$) & \textbf{mALE}~($\downarrow$) \\
    \hline
    SELDNet~\cite{seldnet} & 0.103 $\pm$ 0.002 & 0.501 $\pm$ 0.001 & 0.923 $\pm$ 0.001\\
    SELDNet + mvRGBD & 0.208 $\pm$ 0.001 & 0.635 $\pm$ 0.001 & 0.852 $\pm$ 0.001\\
        \hline
    EIN-v2~\cite{ein_v2} & 0.113 $\pm$ 0.002 & 0.607 $\pm$ 0.001 & 0.878 $\pm$ 0.001\\
    EIN-v2 + mvRGBD & 0.145 $\pm$ 0.001 & 0.687 $\pm$ 0.001 & 0.822 $\pm$ 0.001\\
    \hline
    SALSA~\cite{salsa} & 0.147 $\pm$ 0.002 & 0.722 $\pm$ 0.002 & 0.793 $\pm$ 0.003\\
    SALSA + mvRGBD & 0.289 $\pm$ 0.001 & 0.810 $\pm$ 0.001 & 0.700 $\pm$ 0.002\\
    \hline
    SALSA-Lite~\cite{salsa_lite} & 0.130 $\pm$ 0.010 & 0.712 $\pm$ 0.003 & 0.810 $\pm$ 0.002\\
    SALSA-Lite + mvRGBD & 0.269 $\pm$ 0.003 & 0.792 $\pm$ 0.001 & 0.732 $\pm$ 0.001\\
    \hline
    Sound3DVDet~\cite{He_2024_WACV} & 0.309 $\pm$ 0.010 & 0.998 $\pm$ 0.007 & 0.586 $\pm$ 0.009\\
    Sound3DVDet~\cite{He_2024_WACV} + mvRGBD & 0.378 $\pm$ 0.006 & 0.999 $\pm$ 0.006 & 0.501 $\pm$ 0.005\\
    \hline
    \end{tabular}
    \label{table:supp:add_mvrgbd}
\end{table}

\subsection{More Qualitative Result}

\begin{table*}[t]
    \centering
    \small
    \caption{Quantitative Result w.r.t. Each Sound Source Classes.}
    \vspace{-3mm}
    \begin{tabular}{l|ccc|ccc|ccc}
    \hline
    \multirow{2}{*}{Methods} & \multicolumn{3}{c|}{Telephone} & \multicolumn{3}{c|}{Siren} & \multicolumn{3}{c}{Alarm} \\
    \cline{2-10}
    &AP & AR & ALE &AP & AR & ALE &AP & AR & ALE  \\
    \hline
    SELDNet~\cite{seldnet}       & 0.104 & 0.503 & 0.918  & 0.102 & 0.500 & 0.924 & 0.103 & 0.500 & 0.924 \\
    EIN-v2~\cite{ein_v2}         & 0.112 & 0.606 & 0.881  & 0.114 & 0.608 & 0.881 & 0.112 & 0.607 & 0.879 \\
    SoundDoA~\cite{sounddoa}     & 0.210 & 0.764 & 0.801  & 0.213 & 0.760 & 0.800 & 0.210 & 0.761 & 0.793 \\
    SALSA~\cite{salsa}           & 0.144 & 0.723 & 0.791  & 0.146 & 0.722 & 0.794 & 0.149 & 0.719 & 0.794 \\
    SALSA-Lite~\cite{salsa_lite} & 0.126 & 0.710 & 0.812  & 0.131 & 0.712 & 0.808 & 0.128 & 0.715 & 0.811 \\
    SoundDet~\cite{sounddet}     & 0.119 & 0.670 & 0.820  & 0.120 & 0.675 & 0.823 & 0.117 & 0.672 & 0.825 \\
    Sound3DVDet~\cite{He_2024_WACV} & 0.308 & 0.999 & 0.600 & 0.320 & 0.997 & 0.577 & 0.320 & 0.998 & 0.579 \\
    \hline
    SL3D\_noRGB                  & 0.499 & 0.944 & 0.513 & 0.497 & 0.945 & 0.510 & 0.499 & 0.947 & 0.509 \\
    SL3D\_noDepth                & 0.471 & 0.911 & 0.459 & 0.473 & 0.908 & 0.453 & 0.473 & 0.908 & 0.458 \\
    SL3D\_noCVC                  & 0.500 & 0.948 & 0.391 & 0.504 & 0.949 & 0.387 & 0.497 & 0.945 & 0.391 \\
    SL3D\_noRGBD                 & 0.330 & 0.730 & 0.811 & 0.327 & 0.736 & 0.807 & 0.327 & 0.729 & 0.814 \\
    SL3D\_Res50                  & 0.494 & 0.930 & 0.522 & 0.490 & 0.937 & 0.527 & 0.487 & 0.933 & 0.518 \\
    SL3D\_noDeepSup              & 0.488 & 0.962 & 0.392 & 0.490 & 0.961 & 0.391 & 0.485 & 0.965 & 0.388 \\
    \hline
    Ours \emph{\name} &      \textbf{0.519} & \textbf{0.999} & \textbf{0.317} & \textbf{0.517} & \textbf{0.998} & \textbf{0.323} & \textbf{0.523} & \textbf{0.998} & \textbf{0.324} \\
    \hline
    \hline
        \multirow{2}{*}{Methods} & \multicolumn{3}{c|}{Fireplace} & \multicolumn{3}{c|}{Horn-beeps} & \multicolumn{3}{c}{Overall}\\
    \cline{2-10}
    &AP & AR & ALE &AP & AR & ALE & \textbf{mAP} & \textbf{mAR} & \textbf{mALE} \\
    \hline
    SELDNet~\cite{seldnet}       & 0.101 & 0.499 & 0.923& 0.104 & 0.500 & 0.926 & 0.103 & 0.501 & 0.923\\
    EIN-v2~\cite{ein_v2}         & 0.115 & 0.608 & 0.873 & 0.113 & 0.605 & 0.870 & 0.113 & 0.607 & 0.878\\
    SoundDoA~\cite{sounddoa}     & 0.214 & 0.761 & 0.800 & 0.212 & 0.761 & 0.796 & 0.212 & 0.762 & 0.800\\
    SALSA~\cite{salsa}           & 0.146 & 0.723 & 0.792 & 0.147 & 0.720 & 0.791 & 0.147 & 0.722 & 0.793\\
    SALSA-Lite~\cite{salsa_lite} & 0.131 & 0.714 & 0.810 & 0.130 & 0.710 & 0.809 & 0.130 & 0.712 & 0.810\\
    SoundDet~\cite{sounddet}     & 0.118 & 0.676 & 0.824 & 0.121 & 0.671 & 0.820 & 0.120 & 0.674 & 0.823\\
    Sound3DVDet~\cite{He_2024_WACV} & 0.322 & 0.999 & 0.586 & 0.220 & 0.996 & 0.588 & 0.301 & \cellcolor{secondcolor}0.998 & 0.584 \\
    \hline
    SL3D\_noRGB                  & 0.499 & 0.940 & 0.512 & 0.498 & 0.945 & 0.512 & \cellcolor{thirdcolor}0.498 & 0.944 & 0.510 \\
    SL3D\_noDepth                & 0.471 & 0.910 & 0.457 & 0.473 & 0.911 & 0.459 & 0.472 & 0.910 & 0.457 \\
    SL3D\_noCVC                  & 0.501 & 0.950 & 0.391 & 0.503 & 0.946 & 0.387 & \cellcolor{secondcolor}0.501 & 0.948 & \cellcolor{secondcolor}0.389 \\
    SL3D\_noRGBD                 & 0.330 & 0.734 & 0.809 & 0.331 & 0.732 & 0.813 & 0.328 & 0.732 & 0.810 \\
    SL3D\_Res50                  & 0.488 & 0.929 & 0.522 & 0.489 & 0.933 & 0.519 & 0.488 & 0.932 & 0.520 \\
    SL3D\_noDeepSup              & 0.492 & 0.964 & 0.394 & 0.488 & 0.962 & 0.387 & 0.487 & \cellcolor{thirdcolor}0.960 & \cellcolor{thirdcolor}0.391\\
    \hline
    Ours SoundLoc3D &    \textbf{0.518} & \textbf{0.999} & \textbf{0.315} & \textbf{0.520} & \textbf{0.997} & \textbf{0.317} & \cellcolor{topcolor}\textbf{0.518} & \cellcolor{topcolor}\textbf{0.999} & \cellcolor{topcolor}\textbf{0.320}\\
    \hline
    \end{tabular}
    \label{table:quant_rst_ssclass}
\end{table*}

\begin{table*}[th]
    \centering
    \small
    \caption{Quantitative Result w.r.t. Each Physical Object Class.}
    \vspace{-3mm}
    \begin{tabular}{l|ccc|ccc|ccc}
    \hline
    \multirow{2}{*}{Methods} & \multicolumn{3}{c|}{Table} & \multicolumn{3}{c|}{Ceiling} & \multicolumn{3}{c}{Door} \\
    \cline{2-10}
    &mAP & mAR & mALE &mAP & mAR & mALE &mAP & mAR & mALE \\
    \hline
        SELDNet~\cite{seldnet}       & 0.105 & 0.505 & 0.916  & 0.110 & 0.507 & 0.914 & 0.090 & 0.490 & 0.945 \\
    EIN-v2~\cite{ein_v2}             & 0.114 & 0.608 & 0.870  & 0.117 & 0.610 & 0.865 & 0.109 & 0.598 & 0.870 \\
    SoundDoA~\cite{sounddoa}         & 0.211 & 0.765 & 0.800  & 0.215 & 0.766 & 0.793 & 0.200 & 0.750 & 0.800 \\
    SALSA~\cite{salsa}               & 0.145 & 0.725 & 0.790  & 0.147 & 0.725 & 0.787 & 0.140 & 0.705 & 0.803 \\
    SALSA-Lite~\cite{salsa_lite}     & 0.127 & 0.718 & 0.822  & 0.135 & 0.718 & 0.800 & 0.122 & 0.705 & 0.821 \\
    SoundDet~\cite{sounddet}         & 0.677 & 0.815  & 0.122 & 0.682 & 0.813 & 0.110 & 0.660 & 0.833 & 0.108 \\
    Sound3DVDet~\cite{He_2024_WACV}  & 0.266 & 0.970 & 0.611  & 0.300 & 0.980 & 0.602 & 0.348 & 0.990 & 0.581 \\
    \hline
    SL3D\_noRGB                      & 0.501 & 0.946 & 0.507 & 0.507 & 0.951 &0.501 & 0.477 & 0.926 & 0.520 \\
    SL3D\_noDepth                    & 0.473 & 0.917 & 0.450 & 0.476 & 0.918 & 0.443 & 0.460 & 0.900 & 0.467 \\
    SL3D\_noCVC                      & 0.501 & 0.949 & 0.389 & 0.505 & 0.951 & 0.382 & 0.492 & 0.942 & 0.395 \\
    SL3D\_noRGBD                     & 0.332 & 0.734 & 0.807 & 0.335 & 0.737 & 0.801 & 0.321 & 0.717 & 0.835 \\
    SL3D\_Res50                      & 0.495 & 0.932 & 0.520 & 0.501 & 0.947 & 0.517 & 0.472 & 0.921 & 0.534 \\
    SL3D\_noDeepSup                  & 0.489 & 0.964 & 0.390 & 0.488 & 0.966 & 0.384 & 0.484 & 0.958 & 0.399 \\
    \hline
    SoundLoc3D & \textbf{0.520} & \textbf{0.998} & \textbf{0.318} & \textbf{0.519} & \textbf{0.999} & \textbf{0.318} & \textbf{0.514} & \textbf{0.997} & \textbf{0.327} \\
    \hline
    \hline
        \multirow{2}{*}{Methods}  & \multicolumn{3}{c|}{Chair} & \multicolumn{3}{c|}{Wall} & \multicolumn{3}{c}{Cabinet}\\
    \cline{2-10}
    &mAP & mAR & mALE &mAP & mAR & mALE & mAP & mAR & mALE \\
    \hline
    SELDNet~\cite{seldnet}       & 0.089 & 0.481 & 0.950& 0.108 & 0.505 & 0.919 & 0.091 & 0.497 & 0.926\\
    EIN-v2~\cite{ein_v2}         & 0.100 & 0.578 & 0.894 & 0.117 & 0.610 & 0.868 & 0.103 & 0.589 & 0.898\\
    SoundDoA~\cite{sounddoa}     & 0.200 & 0.750 & 0.812 & 0.215 & 0.767 & 0.780 & 0.210 & 0.757 & 0.802\\
    SALSA~\cite{salsa}           & 0.137 & 0.709 & 0.810 & 0.150 & 0.729 & 0.771 & 0.138 & 0.708 & 0.813\\
    SALSA-Lite~\cite{salsa_lite} & 0.120 & 0.701 & 0.841 & 0.139 & 0.722 & 0.798 & 0.121 & 0.707 & 0.818\\
    SoundDet~\cite{sounddet}     & 0.108 & 0.657 & 0.849 & 0.125 & 0.678 & 0.809 & 0.107 & 0.662 & 0.820\\
    Sound3DVDet~\cite{He_2024_WACV} & 0.220 & 0.923 & 0.613 & 0.294 & 0.990 & 0.588 & 0.300 & 0.975 & 0.579 \\
    \hline
    SL3D\_noRGB                  & 0.470 & 0.920 & 0.544 & 0.504 & 0.950 & 0.500 & 0.480 & 0.927 & 0.543 \\
    SL3D\_noDepth                & 0.461 & 0.900 & 0.469 & 0.483 & 0.923 & 0.430 & 0.462 & 0.901 & 0.469 \\
    SL3D\_noCVC                  & 0.489 & 0.932 & 0.410 & 0.505 & 0.952 & 0.367 & 0.492 & 0.940 & 0.396 \\
    SL3D\_noRGBD                 & 0.312 & 0.712 & 0.819 & 0.339 & 0.741 & 0.802 & 0.317 & 0.729 & 0.828 \\
    SL3D\_Res50                  & 0.477 & 0.912 & 0.539 & 0.498 & 0.948 & 0.530 & 0.480 & 0.928 & 0.528 \\
    SL3D\_noDeepSup              & 0.478 & 0.957 & 0.410 & 0.498 & 0.973 & 0.362 & 0.469 & 0.950 & 0.412\\
    \hline
    SoundLoc3D &  \textbf{0.513} & \textbf{0.998} & \textbf{0.330} & \textbf{0.520} & \textbf{0.999} & \textbf{0.313} & \textbf{0.510} & \textbf{0.997} & \textbf{0.329}\\
    \hline
    \end{tabular}
    \label{table:quant_rst_objclass}
\end{table*}

\begin{figure*}[th]
    \centering
    \includegraphics[width=0.99\linewidth]{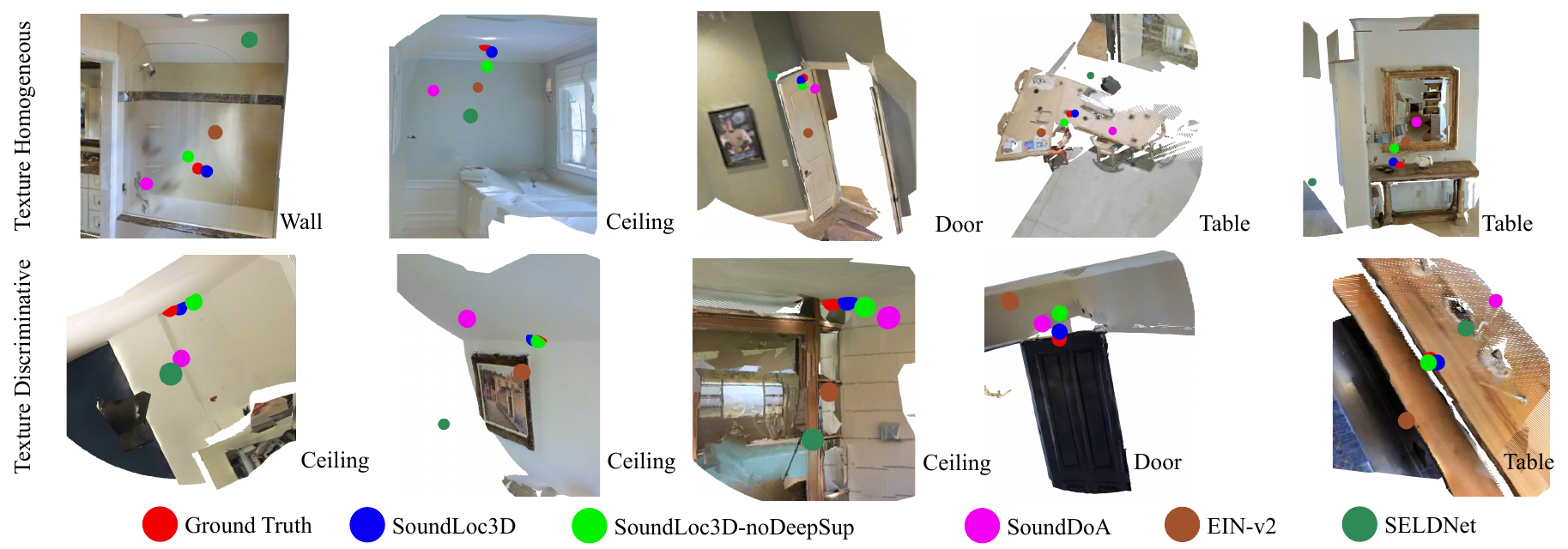}
    \caption{\textbf{More qualitative result}: We visualize the localization result for one sound source in different visual scenes. We also provide the visualization source code and data for more directive visualization.}
    \label{fig:more_qual_vis}
\end{figure*}

We provide more qualitative result visualization in Fig.~\ref{fig:more_qual_vis}. From this figure, we can clearly see that \textit{SoundLoc3D} is capable of accurately detect 3D sound sources under various room scenarios. It is better at handling both texture-homogeneous and texture-discriminative situations.
\end{document}